\documentclass[10pt,conference]{IEEEtran}
\IEEEoverridecommandlockouts
\usepackage{graphicx}
\usepackage{amsmath}
\usepackage{amssymb}
\usepackage{algorithm}
\usepackage{algorithmic}
\usepackage[T1]{fontenc}
\usepackage{color}
\usepackage{subfigure}
\usepackage{booktabs}
\usepackage{cases}
\usepackage{cite}
\usepackage{footmisc}
\usepackage{CJK}
\usepackage{type1cm}
\usepackage{times}
\usepackage{multicol}
\usepackage{multirow}
\usepackage{setspace}
\usepackage{array}
\usepackage{adjustbox}
\usepackage{tabularx}
\usepackage{booktabs}

\usepackage[top=0.7in, bottom=0.94in, left=0.61in, right=0.61in]{geometry}

\usepackage{fancyhdr}
\fancypagestyle{firststyle}{\fancyhf{}
	\fancyhead[L]{M. Ying, D. Shakya, H. Poddar, and T. S. Rappaport, ``Waste Factor: A New Metric for Evaluating Power Efficiency in any Cascade", in\textit{ GLOBECOM 2023 - 2023 IEEE Global Communications Conference, }Kuala Lumpur, Malaysia, Dec. 2023, pp. 1--6.}
}

\begin{document}
	
\title{Waste Factor: A New Metric for Evaluating Power Efficiency in any Cascade}

\author{
	\IEEEauthorblockN{Mingjun Ying, Dipankar Shakya, Hitesh Poddar, and Theodore S. Rappaport}
\IEEEauthorblockA{ NYU WIRELESS, Tandon School of Engineering, New York University, Brooklyn, NY, 11201\\
 \{yingmingjun, dshakya, hiteshp, tsr\}@nyu.edu}
 \thanks{This research is supported by the New York University (NYU) WIRELESS
 Industrial Affiliates Program, and NYU Tandon School of Engineering graduate fellowship.}
 }

\maketitle
% \linespread{0.95}
\newcommand\blfootnote[1]{%
	\begingroup
	\renewcommand\thefootnote{}\footnote{#1}%
	\addtocounter{footnote}{-1}%
	\endgroup
}

	% \blfootnote{\noindent{This research is supported by the New York University (NYU) WIRELESS
	% Industrial Affiliates Program, and NYU Tandon School of Engineering graduate fellowship.}}

\thispagestyle{firststyle}
\begin{abstract}
	In this paper, we expand upon a new metric called the Waste Factor (\textit{W}), a mathematical framework used to evaluate power efficiency in cascaded communication systems, by accounting for power wasted in individual components along a cascade. We show that the derivation of the Waste Factor, a unifying metric for defining wasted power along the signal path of any cascade, is similar to the mathematical approach used by H. Friis in 1944 to develop the Noise Factor (\textit{F}), which has since served as a unifying metric for quantifying additive noise power in a cascade. Furthermore, the mathematical formulation of \textit{W} can be utilized in artificial intelligence (AI) and machine learning (ML) design and control for enhanced power efficiency. We consider the power usage effectiveness (PUE), which is a widely used energy efficiency metric for data centers, to evaluate \textit{W} for the data center as a whole. The use of \textit{W} allows easy comparison of power efficiency between data centers and their components. Our study further explores how insertion loss of components in a cascaded communication system influences \textit{W} at 28 GHz and 142 GHz along with the data rate performance, evaluated using the consumption efficiency factor (CEF). We observe CEF's marked sensitivity, particularly to phase shifter insertion loss changes. Notably, CEF variations are more prominent in uplink transmissions, whereas downlink transmissions offer relative CEF stability. Our exploration also covers the effects of varying User Equipment (UE) and Base Station (BS) deployment density on CEF in cellular networks. This work underscores the enhanced energy efficiency at 142 GHz, compared to 28 GHz, as UE and BS numbers escalate.
	\end{abstract}
	

\section{Introduction}

As the evolution of telecommunications progresses from 5G towards 6G, there is an escalating demand for energy efficiency in both wired and wireless communication systems. Currently, these systems account for approximately 2-3\% of the global energy demand, a figure projected to exceed 20\% by 2030 \cite{a1,a10}. This rise is primarily due to the forthcoming 5G and 6G networks and edge computing, which promise ultra-wide bandwidths and increased data rates. These advancements, however, intensify the challenge of managing energy efficiency, especially within the context of resource-limited Internet of Things (IoT) \cite{a2,a3}. The critical need to reduce energy consumption in wireless networks also stems from the urgency to limit greenhouse gas emissions and mitigate climate change as machine learning (ML) and artificial intelligence (AI) threaten to expand power consumption from their computational burden.

Wasted power is becoming a significant enemy of the planet, echoing a parallel from the past. Over 80 years ago, noise was the primary adversary to wireless communication. It was H. Friis who developed the Noise Factor and Noise Figure in dB, a unifying metric to assess additive noise power in a cascade \cite{a12}. Today, Waste Factor ($W$) emerges as an analogous tool to assess wasted power in a cascade system, the current adversary to our environment and planet's energy resources.

The advancements in massive MIMO, network slicing, renewable energy-powered BS, and energy harvesting technologies have significantly contributed to reducing energy consumption in 5G networks \cite{m1, m2}. Furthermore, artificial intelligence (AI) and machine learning (ML) techniques, including reinforcement learning, may be instrumental in maintaining a balance between quality of service (QoS) and energy consumption \cite{m3}. Despite these strides, a glaring gap persists in existing research and design - the lack of a comprehensive theoretical framework to measure and compare power efficiency across diverse wireless system architectures \cite{a6,a9,a8}.

Waste Factor $W$ and Consumption Efficiency Factor (CEF) \cite{a8, a9, a10,x1} aim to fill this gap. By providing a standardized metric for comparing power consumption and energy efficiency across various system designs, W and CEF can guide engineers and product designers towards more sustainable and energy-efficient solutions, towards minimal power consumption in future wireless network design, particularly those operating at sub-THz frequencies \cite{a10}. Simulations in \cite{a10} also demonstrate that reducing cell size and increasing carrier frequency and bandwidth lead to lower energy expends per bit, confirming the relevance and utility of $W$ in achieving energy-efficient designs. In this paper, we further extend and apply the theoretical constructs of W and CEF, making the following contributions:

\begin{itemize}
\item While the concept of Waste Factor was first introduced in \cite{a10} and \cite{x1}, we provide here for the first time a detailed mathematical derivation of $W$ as well as intuitive analogies to Noise Factor based on H. Friis original mathematical derivations in \cite{a12}. 
\item We extend Waste Factor to generalized communication systems, using the case of energy consumption in data centers as a primary example. We also illustrate the effectiveness of $W$ through an example comparing the power waste of two data centers. 
\item We explore the impact of varying component efficiency for a TX and RX cascade, particularly the phase shifter's insertion loss on CEF at 28 GHz and 142 GHz. 
\item We analyze the influence of user equipment (UE) and base station (BS) geographic density on network CEF.
\item While not covered here, we note that $W$ could be used in AI/ML applications to optimize power effiency.

\end{itemize}

The structure of this paper is as follows. Section \uppercase\expandafter{\romannumeral2} derives the Waste Factor, drawing an analogy between its mathematical form and that of the Noise Factor. Section \uppercase\expandafter{\romannumeral3} uses $W$ to analyze data centers. In Section \uppercase\expandafter{\romannumeral4}, we apply $W$ to analyze the energy efficeiency of future communication systems. Section \uppercase\expandafter{\romannumeral5} discusses potential future research directions using Waste Factor. Finally, Section \uppercase\expandafter{\romannumeral6} concludes the paper.

\section{Introduction to F and W}
In this section we show the duality of two parameters, $F$ and $W$, that can be used to evaluate noise and power waste of communication systems, respectively. 
\vspace{-3pt}
\subsection{Noise Factor}
Noise factor ($F$) defines to the degradation of signal-to-noise ratio (SNR) in a cascade. Specifically, the noise factor $F$ is defined as the ratio of the input SNR to output SNR, which is expressed as $F = {SNR_{i}}/{SNR_{o}}$. $F$ in dB is the noise figure (NF) and a value of 0 dB indicates no added noise and no degradation in SNR along a device or cascade. Friis's formula iswidely used to calculate the overall $F$ of cascaded devices, where each device has its own individual $F$ and power gain, $G$. Once the total $F$ is calculated, it can be used to determine the overall NF of the entire cascade. Based on \cite{a12}, $F$ for the cascaded system is
\vspace{-3pt}  
\begin{equation}\label{b1}
		F=F_{1}+\frac{\left(F_{2}-1\right)}{G_1}
		+\frac{\left(F_{3}-1\right)}{G_1 G_{2}}
		 +\ldots+\frac{\left(F_N-1\right)}{\prod_{i=1}^{N-1} G_i},
		 \vspace{-4pt}
\end{equation}
where $F_{i}$ represents the noise factor of the i-th device, and $G_{i}$ represents its power gain (linear, not in dB).
\vspace{-3pt}

\subsection{Waste Factor ($W$)}
The Waste Factor $W$ characterizes power efficiency of a cascaded system by considering the power wasted by components along a cascade. Akin to $F$, the power wasted by a device/cascade can also be examined by observing the progressive power waste, based on the output power at each stage, as the signal propagates down the cascade. Such formulation provides an intuitive way to understand power waste at each stage of the cascade and allows $W$ to compare the power efficiency of two devices/systems through wasted power. For analysis, the power consumed ($P_{consumed}$) is split into three principal components \cite{ a9, a10, x1}: 
\begin{itemize}
	\item Signal path power ($P_{signal}$): Power delivered to the device/cascade output (e.g., power amplifier output to matched load). 
	\item Non-signal path power ($P_{non-signal}$): Power consumed by devices on the path to facilitate signal transfer in the cascade (e.g., standby power drawn by an amplifier).
	\item Non-path power ($P_{non-path}$): Power consumption of components that do not contribute to the signal and are not along the cascade (e.g., oscillators, displays, etc.). 
\end{itemize}

Thus, we have
\vspace{-3pt}
\begin{equation}\label{e1}
	P_{consumed}=(P_{signal}+P_{non-signal})+P_{non-path}.
\end{equation}

Fundamentally, $W$ is defined as the ratio of power consumed by the signal path components ($P_{consumed,\, path}= P_{signal}+P_{non-signal}$) to the useful signal power ($P_{signal,out}$) delivered along the cascaded communication system ($W = P_{consumed,\, path}/P_{signal,out}$)\cite{a10}. Since $W$ is based on the useful signal power output, it is referred to the output. 
\vspace{-10pt}
\begin{figure}[htbp]
	\centering
	\includegraphics[width=2.8in]{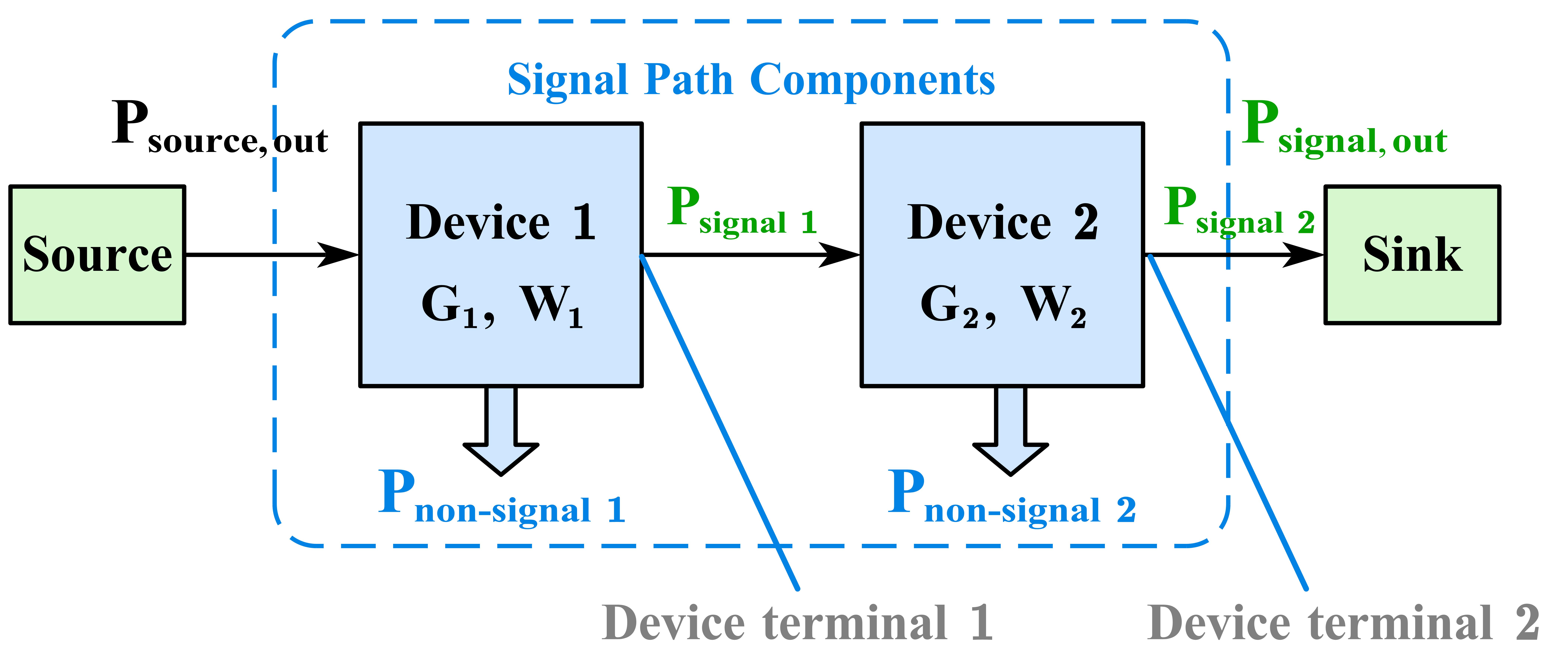}
	\vspace{-10pt}
	\caption{A general cascade communication system composed of two devices.}
	\label{figa1}
	\vspace{-8pt}
\end{figure}

The formulation of $W$ for a cascaded system is illustrated through a simple cascade of two devices in Fig. \ref{figa1}, neglecting $P_{non-path}$ from auxiliary components. Here, we define  
% The formulation of $W$ for a cascaded system is illustrated through a simple cascade of two devices in Fig. \ref{figa1}, neglecting $P_{non-path}$ from auxiliary components. Here, we define  
\vspace{-3pt}
\begin{equation}\label{d1}
	\begin{aligned}
		P_{consumed}&=W \times P_{signal,out}.\\
	\end{aligned}
	\vspace{-3pt}
\end{equation}

Now, we can define the power consumed at device 1's terminal as:
\vspace{-3pt}
\begin{equation}\label{d2}
	P_{consumed,D1}=W_{1}  P_{signal\,1},
	\vspace{-3pt}
\end{equation}
Here, $P_{consumed,D1}$ denotes the total power consumption at the output terminal of device 1. This comprises both the useful signal power transmitted to the subsequent device and the power wasted by device 1 itself. When $P_{source,out}$ is subtracted, we arrive at the standalone power consumption of device 1.
\vspace{-3pt}
\begin{equation}\label{d3}
	P_{consumed\,1}=W_{1}  P_{signal\,1}-P_{source,out},
	\vspace{-3pt}
\end{equation}
similar to that, the power consumption of device 2 alone is
\vspace{-3pt}
\begin{equation}\label{d4}
	P_{consumed\,2}=W_{2}  P_{signal\,2}-P_{signal\,1}.
	\vspace{-3pt}
\end{equation}

Intuitively, the total power consumed can be expressed as in \eqref{d1_1}. This total power consumption is the sum of power consumed by each device and the power input to the system.
\vspace{-3pt}
\begin{equation}\label{d1_1}
	\begin{aligned}
		P_{consumed}& =P_{consumed\,1}+P_{consumed\,2}+P_{source,out}.
	\end{aligned}
	\vspace{-3pt}
\end{equation}

Also, we know that the output of the system is
\vspace{-3pt}
\begin{equation}\label{d5}
	P_{signal,out}=P_{signal\,2}=G_{2}P_{signal\,1},
	\vspace{-3pt}
\end{equation}
based on (\ref{d2})-(\ref{d5}), we have
\vspace{-3pt}
\begin{equation}\label{d6}
	\begin{aligned}
		P_{consumed}&=W_{2}P_{signal\,2}+(W_{1}-1)P_{signal\,1}\\
		&=(W_{2}+\frac{(W_{1}-1)}{G_{2}})P_{signal,out}.
	\end{aligned}
	\vspace{-3pt}
\end{equation}

Since (\ref{d6}) is equal to (\ref{d1}), then we have the power waste factor for the cascaded system
\vspace{-3pt} 
\begin{equation}\label{d7}
	W=(W_{2}+\frac{(W_{1}-1)}{G_{2}}).
	\vspace{-3pt}
\end{equation}

Based on \eqref{d7}, $W$ for a cascaded system with $N$ devices can be generalized to \eqref{f7}. It is noteworthy that the mathematics below bear a striking similarity to the Noise Factor in (\ref{b1}) \cite{a12}.
\vspace{-3pt}
\begin{equation}\label{f7}
	W=\{{{W}_{N}}+\frac{\left(W_{N-1}-1\right)}{G_N}
	+\frac{\left(W_{N-2}-1\right)}{G_N G_{N-1}}
	+\ldots+\frac{\left({W_1}-1\right)}{\prod_{i=2}^N G_i}\}.
	% \vspace{-3pt}
\end{equation}

\vspace{-10pt}

\subsection{Analogies between $F$ and $W$}    

\begin{table*}[!t]
	\small
	\caption{Comparison between Waste Factor and Noise Factor}
	\vspace{-3pt}
	\label{tab:comparison}
	\centering
	\renewcommand{\arraystretch}{1.1}
	\begin{tabular}{|m{3.5cm}|p{6.7cm}|p{6.7cm}|}
	\hline
	\textbf{Aspect} & \textbf{Waste Factor} & \textbf{Noise Factor} \\
	\hline
	\textbf{Definition} & Indicates the amount of power wasted in a device or cascade when referred to the output.
	& Indicates the amount of additive noise power from device or cascade referred to the input.
	\\
	\hline
	\textbf{Port of Reference} & Referenced to the output of a device or cascade. & Referenced to the input of a device or cascade. \\
	\hline
	\textbf{Interpretation of Value} & $W = 1$ (0 dB) indicates all power consumed by a device or cascade is delivered as the output and no power is wasted. Higher $W$ means higher power waste. & $F = 1$ indicates there is no additive noise in a device or cascade and there is no degradation in SNR. Higher $F$ means higher SNR degradation. \\
	\hline
	\textbf{Utility} & Used to analyze the power consumed and wasted by wireless networks and devices. & Widely used in industry to analyze the additive noise in communication systems and devices.\\
	\hline
	\end{tabular}
	\vspace{-15pt}
	\end{table*}

The analogous mathematical formulation of $F$ and $W$ is immediately visible from \eqref{b1} and \eqref{f7}. There are, however, important characteristics of each metric to keep in mind. 

As the noise figure is a measure of the degradation of the SNR caused by the components in a cascaded system, it quantifies the amount of noise added to the signal at the input of the cascade. Therefore, $F$ is referred to the cascade input and \eqref{b1} increases from device 1 to N (source to sink). On the other hand, $W$ is a measure of the power efficiency of a cascaded system. It quantifies the amount of power consumed by the cascade to transmit or receive a signal. Since $W$  is related to the power consumed by the cascade, it is referred to the cascade output and \eqref{f7} increases from device N to 1 (sink to source). 

A higher $W$ intuitively signifies more power wasted. The value of $W$ is always equal to or greater than 1, with $W = 1$ signifying that all power supplied to a cascaded component or network is fully utilized in the signal output (optimal, no power wasted). Conversely, $W \to \infty$ indicates that no power is contributed to the signal output, and all power is squandered (e.g., a perfect dummy load or an entirely lossy channel). The aspects of the $F$ and $W$ in communication systems are comprehensively summarized in Table~\ref{tab:comparison}.

In conclusion, both $F$ and $W$ are useful in the analysis of communication systems. $F$ is a well-established metric that provides a measure of the degradation of the signal-to-noise ratio, and $W$ is a new metric that provides a measure of the power efficiency of a system. Both metrics are important for communication systems. With the increasing importance of energy efficiency in the industry, $W$ can become a vital metric for enabling green communications. 
\vspace{-3pt}
\section{Adoption of Waste Factor}
In communication systems, effective use of $W$ is vital for more energy-efficient solutions. Specifically, the potential of $W$ to apply across communication systems and shed light on energy consumption needs exploration. We extend the theoretical framework to all data systems, including data centers, analyzing superposition of power (\ref{e1}) to understand energy consumption, akin to the consumption factor analysis in \cite{a9,a8}. This method helps differentiate power for information conveyance from other processes off the cascaded path.
\vspace{-3pt}
\begin{equation}\label{e7}
	\begin{aligned}
		W &= {W_{sink}}+\frac{1}{{{G}_{RX}}}\left( \frac{1}{{{G}_{channel\text{ }}}}-1 \right)\\
		\vspace{-2pt}
		&+\frac{1}{{{G}_{RX}}{{G}_{channel\text{ }}}}\left( {W_{source}}-1 \right).
	\end{aligned}
	\vspace{-4pt}
\end{equation}

Using $W$, we aim to measure power wastage in devices on a cascaded signal path, assisting engineers in identifying power wastage hotspots for improved efficiency, potentially with AI/ML. Section III examines the use of waste factor concepts for "ancillary functions" in data centers, like cooling, lighting, and non-path components. Based on \eqref{f7}, we generalize $W$ in \eqref{e7} for wireless systems with a source, sink, and communication channel, as seen in \cite{a9}.
% In the realm of communication systems, effective adoption of $W$ is crucial in order to foster more energy-efficient solutions. In particular, the possibility that $W$ may be applicable to all types of communication systems and offer new insights into energy consumption warrant further exploration. In this paper, we expand the existing theoretical framework to encompass all data communication systems, including data centers, by considering superposition of power (\ref{e1}) as a means of dissecting energy consumption. The analysis resembles the consumption factor analysis in \cite{a9,a8}. This approach allows one to distinguish between the power required for conveying information and that consumed by other ancillary processes off the cascaded path.

% By employing $W$, we aim to quantify the power wasted by all devices on a signal path in a cascade, which in turn enables engineers to pinpoint areas where power is being squandered and implement measures to curtail power waste (perhaps someday with AI/ML). Furthermore, Section III explores the applicability of similar waste factor concepts to govern the "other ancillary things" in data centers and other power consumers, such as cooling systems, lighting, disks, interfaces, and other non-path components. Based on \eqref{f7}, we extend $W$ to a generalized wireless communication system with an information source, sink, and a channel for communicating a signal between the source and sink, we have \cite{a9}:

\subsection{Generalized Waste Factor for data center}
\vspace{-3pt}
A generalized formulation of $W$ can be implemented by taking a data center as an example. The power usage effectiveness (PUE) is a measure of energy efficiency in data centers, which is the ratio between the amount of energy consumed by computing equipment for data operations to the total overhead energy used for supporting equipment, including cooling \cite{a24}. 
\vspace{-3pt}
\begin{equation}
	{PUE}=\frac{\text{Computing equipment energy}}{\text{Auxiliary equipment energy}}.
	\vspace{-3pt}
\end{equation}
% \vspace{-2pt}
Using \eqref{e7} we elucidate our approach to extend the existing mathematical framework for $W$ to all energy-consuming systems within the communication process, ultimately establishing a generalized waste factor that serves as a metric for system energy efficiency. By doing so, we strive to enhance the applicability of $W$ to a wide range of power consumers, including data centers and other complex systems, in order to quantify energy efficiency and reduce overall power consumption. Here we study a data center as an example to explain how $W$ can be applied to a generalized system. 

In the process of data transmission within a data center, the major power consumption is attributed to servers, network switches and computing equipment, while other power consumption is associated with cooling systems, power distribution units (PDUs), and other auxiliary equipment. According to Barroso and Hölzle's findings \cite{a23}, server and networking equipment account for about 60-70\% of the overall power consumption in a data center. Cooling systems contribute about 30-40\% of the total power consumption, while the remainder is consumed by PDUs and other auxiliary equipment.

Total data center power consumption can be modeled as:
\vspace{-3pt}
\begin{equation}\label{b2}
	{{P}_{consumed}}={{P}_{info}}+{{{P}_{non-info}}}+{{{P}_{aux}}},
	\vspace{-3pt}
\end{equation}
where ${P}_{info}$ is the sum of all powers of each component (e.g. routers, switches, processors, and other network equipments that carry or store information) that has been used for carrying information in the system, ${P}_{non-info}$ is the power used by the other components but not directly involved in data transmission (e.g. servers, storage devices, firewalls), and ${P}_{aux}$ is the power used by the cooling systems, PDUs, and other auxiliary equipment apart from the data transmission and access/storage. Based on \cite{a24}, the PUE is a common measure of energy efficiency in data centers. PUE assesses the ratio between the amount of energy consumed by computing equipments and the total overhead energy used for supporting equipments, including cooling. Using (\ref{b2}), we notice that
\vspace{-3pt}
\begin{equation}\label{b3}
	{PUE}=\frac{{{P}_{info}}+{{{P}_{non-info}}}+ P_{aux}}{{P}_{info}+{P}_{non-info}}.
	\vspace{-3pt}
\end{equation}

Let us consier the power efficiency for all data transmissions in the data center as $\eta$
\vspace{-3pt}
\begin{equation}\label{b4}
	{\eta}=\frac{{P}_{info}}{{P}_{info}+P_{non-info}}.
	\vspace{-3pt}
\end{equation}

The power consumption for data processing can be derived considering the data center as a single component
\vspace{-3pt}
\begin{equation}\label{b5}
	{P}_{consumed}={P}_{info} \overline{W}+{{P}_{aux}},
	\vspace{-3pt}
\end{equation}
where $\overline{W}$ represents the Waste Factor for the data center, and using (\ref{b3}), (\ref{b4}), and (\ref{b5}), we find
\vspace{-3pt}
\begin{equation}\label{b6}
	\overline{W}={\eta}^{-1}=\frac{P_{aux}}{P_{info}{(PUE-1)}}.
	\vspace{-3pt}
\end{equation}

Based on (\ref{d7}), here we consider a communication scenario between two data centers. Also, $W$ for the communication channel (e.g. cable or connecgtor loss) is obtained by treating it as a passive attenuator with less than unity gain \cite{a10,x1}, then we determine the $W$ as
\vspace{-3pt}
\begin{equation}\label{c7}
	\begin{aligned}
		W &= \overline{W}_{sink}+\frac{1}{{{G}_{RX}}}\left( \frac{1}{{{G}_{channel\text{ }}}}-1 \right)\\
		&+\frac{1}{{{G}_{RX}}{{G}_{channel\text{ }}}}\left( \overline{W}_{source}-1 \right).
	\end{aligned}
	% \vspace{-4pt}
\end{equation}

\subsection{Comparison of data centers power waste using $W$}

Two data centers: Data Center A and Data Center B, each with different power consumption levels. Despite its wide adoption, the PUE cannot serve as a universal standard for energy efficiency comparison. PUE measures a data center's energy efficiency by comparing total facility power to auxiliary equipment power. However, it doesn't distinguish between power used for actual data transmission and other uses. When two data centers have the same PUE, we can't tell which is more efficient at data transmission. However, $W$ provides a more detailed insight, differentiating between power that carries information and power that doesn't. So, for a nuanced energy efficiency comparison, especially between data centers with identical PUEs, incorporating an additional metric like $W$ is essential. To make the point that $W$ gives more insight than PUE, we consider two data centes with equal PUE but different architecures. For Data Center A, the power allocation is as follows: \(P_{{{infoA}}} = 140\) kWh for information transmission, \(P_{{{non-infoA}}} = 40\) kWh for non-data transmission components, and \(P_{{{auxA}}} = 150\) kWh for auxiliary equipment. In comparison, Data Center B allocates \(P_{{{infoB}}} = 60\) kWh of power for information transmission, \(P_{{{non-infoB}}} = 30\) kWh for non-data transmission components, and \(P_{{{auxB}}} = 75\) kWh for auxiliary equipment.

Here, we assume that Data Center A is a larger facility with more equipment, hence a higher total energy consumption. Conversely, Data Center B is smaller and uses less total energy. If we were to simply compare their total energy use, it might seem that Data Center B is more efficient. But the Waste Factor paints a different picture. First, we calculate the power efficiency for data transmission in \eqref{b4} computing equipment (\(\eta\)) for both data centers:
\vspace{-3pt}
\begin{equation*}\label{q1}
	\eta_A = \frac{P_{{infoA}}}{P_{{infoA}} + P_{{non-infoA}}}=\frac{140}{140 + 40} \approx 0.778,
\end{equation*}
\vspace{-2pt}
\begin{equation*}\label{q2}
	\eta_B = \frac{P_{{infoB}}}{P_{{infoB}} + P_{{non-infoB}}}=\frac{60}{60 + 30} \approx 0.667,
\end{equation*}
where $\eta_A$ and $\eta_B$ denote the power efficiency of overall data transmission in Data Center A and Data Center B, respectively. Next, we calculate the PUE for both data centers using \eqref{b3}:
\vspace{-3pt}
\begin{equation*}\label{q3}
	PUE_A = \frac{{{P}_{infoA}}+{{{P}_{non-infoA}}}+ P_{auxA}}{{P}_{infoA}+{P}_{non-infoA}}  \approx 1.833,
\end{equation*}
\vspace{-2pt}
\begin{equation*}\label{q4}
	PUE_B = \frac{{{P}_{infoB}}+{{{P}_{non-infoB}}}+ P_{auxB}}{{P}_{infoB}+{P}_{non-infoB}} \approx 1.833,
\end{equation*}
where $PUE_A$ and $PUE_B$ denote the PUE of Data Center A and Data Center B, respectively. Finally, the Waste Factor (\(\overline{W}\)) for both data centers is found using \eqref{b6}:
\vspace{-3pt}
\begin{equation*}\label{q5}
	\overline{W}_A =\frac{P_{auxA}}{P_{infoA}{(PUE_{A}-1)}} =\frac{150}{140 \times 0.833} \approx 1.286,
\end{equation*}
\vspace{-3pt}
\begin{equation*}\label{q6}
	\overline{W}_B = \frac{P_{auxB}}{P_{infoB}{(PUE_{B}-1)}} =\frac{75}{60 \times 0.833} \approx 1.5,
	\vspace{-3pt}
\end{equation*}
\vspace{-10pt}

The Waste Factors of both data centers ($\overline{W}_A$ and $\overline{W}_B$) allow us to identify the more energy-efficient one, Data Center A, is more energy efficient due to its lower Waste Factor ($\overline{W}_A<\overline{W}_B$). The PUE metric, despite being a common measure, only accounts for the energy consumed by computing and supporting equipment in the whole, neglecting variations in operational conditions, equipment type, and workload characteristics. This necessitates more comprehensive metrics, such as the Waste Factor. By introducing the generalized \(W\) as an energy efficiency metric, we can calculate the \(W\) for complex systems like data centers using the proposed formula (\ref{c7}). This approach extends the applicability of the Waste Factor to a variety of power consumers, including data centers and other systems with significant energy consumption.
\vspace{-3pt}
\section{Evaluating performance with CEF}
\begin{figure}[!t]
	\vspace{-3pt}
	\centering
	\includegraphics[width=0.38\textwidth]{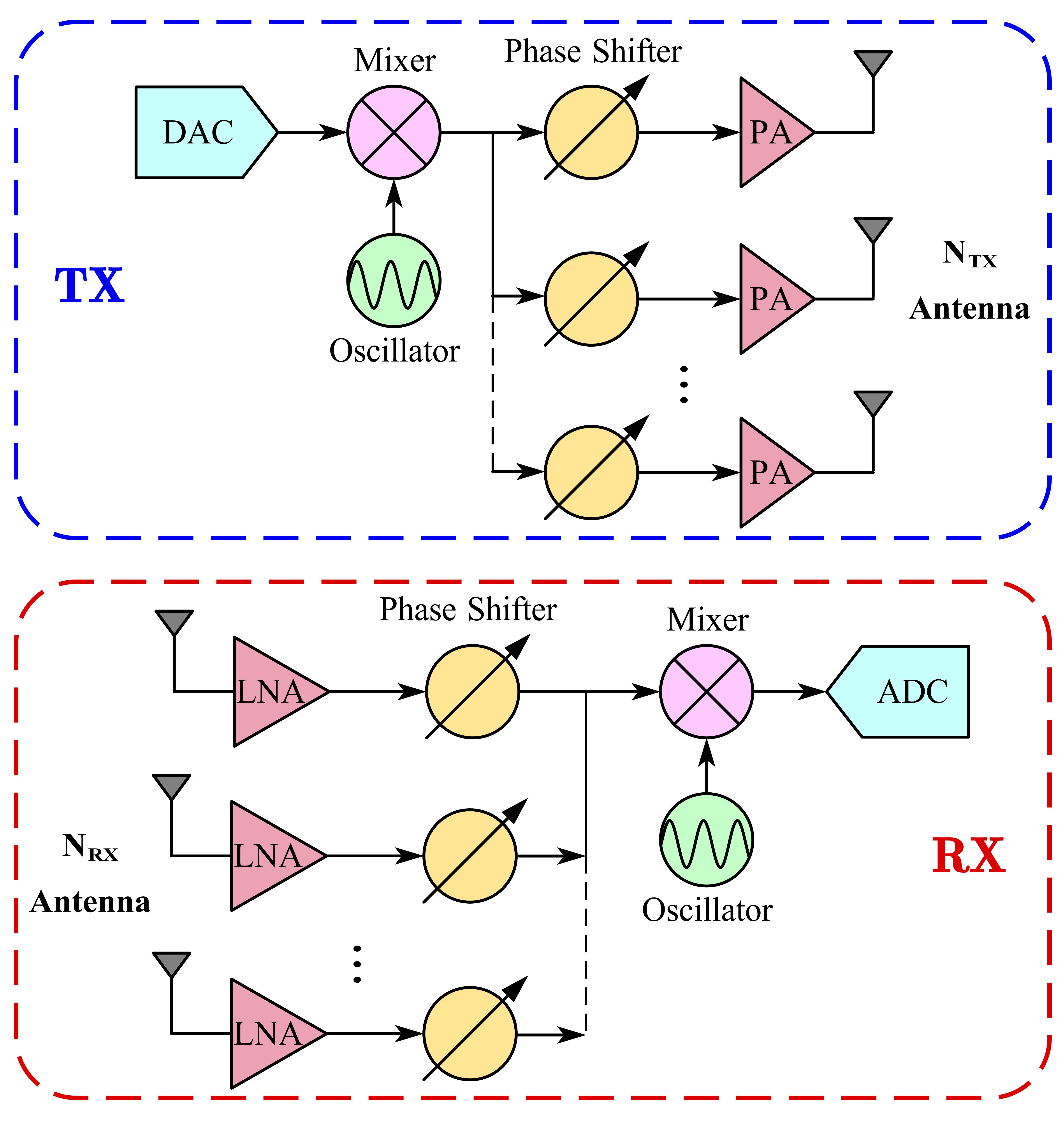}
	\vspace{-9pt}
	\caption{Architecture of the TX and RX considered for power analysis.}
	\label{figa7}
	\vspace{-15pt}
\end{figure}

\vspace{-3pt}
The CEF is defined as the maximum data rate delivered by the communication system to the total power consumed. Using $W$, the CEF can be derived as \cite{a9,a10}:
\vspace{-3pt}
\begin{equation}\label{s1}
	\text{CEF} = \frac{R_{max}}{W \times P_{signal,out}},
	\vspace{-3pt}
\end{equation}
where $R$ is the data rate in bps, and $R_{max}$
is the maximum data rate supported by the communication system. To analyze the impact of component efficiecy on $W$, and ultimately CEF, here we assume a transceiver and receiver structure illustrated by Fig. \ref{figa7} with analog beamforming at both TX and RX. The simulation is conducted for mmWave (28 GHz) and sub-THz (142 GHz) wireless communication systems, and the detailed simulation parameters for the comparison of two communication systems are summarized in Table \ref{tabc}, and the selection of parameters from the table as simulation parameters is mainly based on \cite{a10}. In the following simulations, we employ the CEF owing to its comprehensive encapsulation of system performance, encompassing aspects of both power consumption and data rate. This metric not only assists in the optimization of system efficiency, but also affords a convenient benchmark for comparison across disparate systems or configurations. 

We emphasize here that the low-noise amplifier (LNA) is not considered as an on-path component for power consumption analysis. The reason being that the DC power consumed by the LNA is independent of the input signal power. In other words, the gain and efficiency of the LNA remain relatively constant within its linear operating range, and do not vary significantly with input power. Since the input power to the LNA is typically weak, coming from the RX antenna, the LNA's gain of over 30 dB amplifies the signal to a level that can be processed by the subsequent circuitry. In our analysis, we classify the power consumed by the LNA differently. Since its consumption is consistent, we categorize it as non-path power when calculating the system's total power consumption. Separately, for an accurate representation of its operational efficiency, we factor in the LNA's Figure of Merit (FoM). This approach allows us to estimate the system's overall power consumption accurately, keeping our model reasonably straightforward.
% \vspace{-5pt}
\begin{table*}[!t]
	\small
	\setlength{\tabcolsep}{10pt} % 调整列间距
	\renewcommand{\arraystretch}{0.8} % 调整行间距
	\caption{Simulation Parameters Comparison of mmWave and Sub-THz systems}
	\vspace{-3pt}
	\label{tabc}
	\centering
	\begin{tabular}{@{}p{6cm}p{3.5cm}p{3.5cm}c@{}}
	\toprule
	\textbf{Parameter} & \textbf{mmWave (28 GHz)} & \textbf{Sub-THz (142 GHz)} & \textbf{Units}\\
	\midrule
	Bandwidth & 400 & 4000 & MHz \\
	Antenna aperture area - BS & 0.5 & 0.5 & m$^2$ \\
	Antenna aperture area - UE & 0.0005 & 0.0005 & m$^2$ \\
	Antenna gain - BS & 45.2 & 59.1 & dBi \\
	Antenna gain - UE & 15.2 & 29.1 & dBi \\
	Path loss exponent - LOS\,|\,NLOS & 2.0\,|\,3.2 & 2.0\,|\,3.2 & - \\
	Number of BS antenna elements & 1024 & 4096 & - \\
	Number of UE antenna elements & 8 & 64 & - \\
	Low-noise amplifiers (LNA) figure of merit & 24.83 & 8.33 & mW$^{-1}$ \\
	LNA gain & 20 & 20 & dB \\
	Mixer insertion loss & 6 & 6 & dB \\
	Phase shifter insertion loss & 10 & 10 & dB \\
	Local oscillator power & 10 & 19.9 & dBm \\
	TX power amplifier (PA) efficiency & 28 & 20.8 & \% \\
	Antenna efficiency & 0.6 & 0.6 & - \\
	Cooling overhead at the BS & 20 & 20 & \% \\
	UE screen power consumption & 500 & 500 & mW \\
	\bottomrule
	\end{tabular}
	\vspace{-5pt}
	\end{table*}

	\vspace{-3pt}
\begin{figure*}[!t]
    \centering
    \subfigure[The effect of PS insertion loss (dB) on CEF.]{\includegraphics[width=0.31\textwidth]{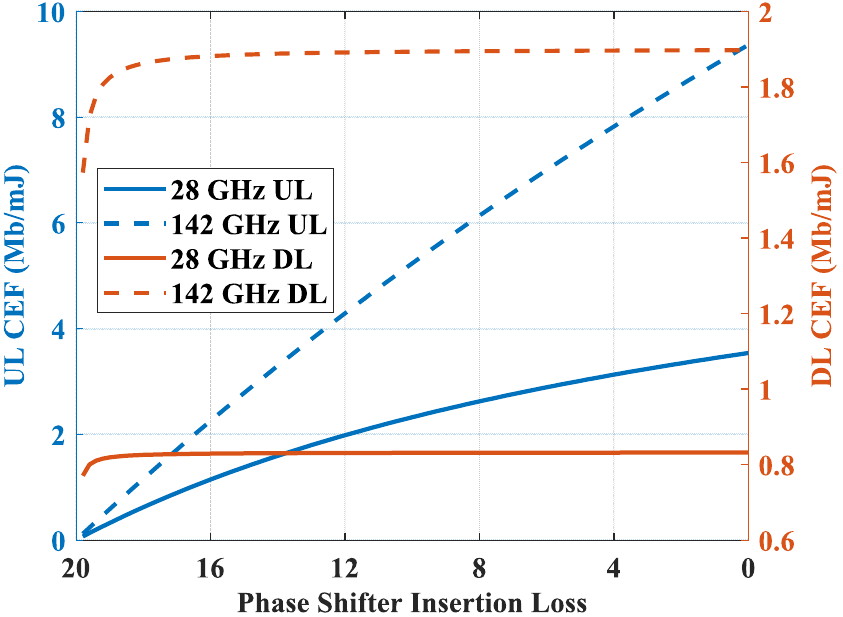}\label{figa4}}
    \hfil
    \subfigure[The effect of the UE number on the CEF.]{\includegraphics[width=0.31\textwidth]{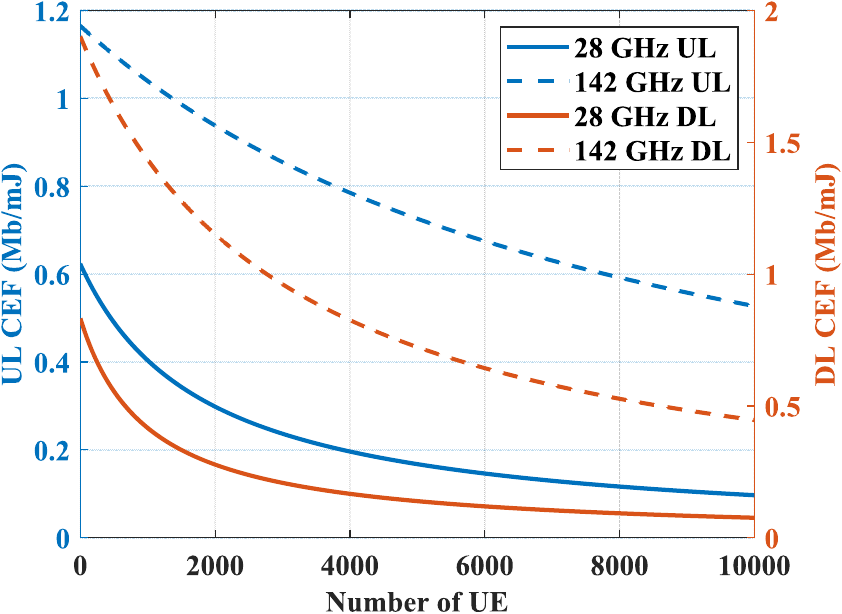}\label{figa5}}
    \hfil
    \subfigure[The effect of the BS number on the CEF.]{\includegraphics[width=0.31\textwidth]{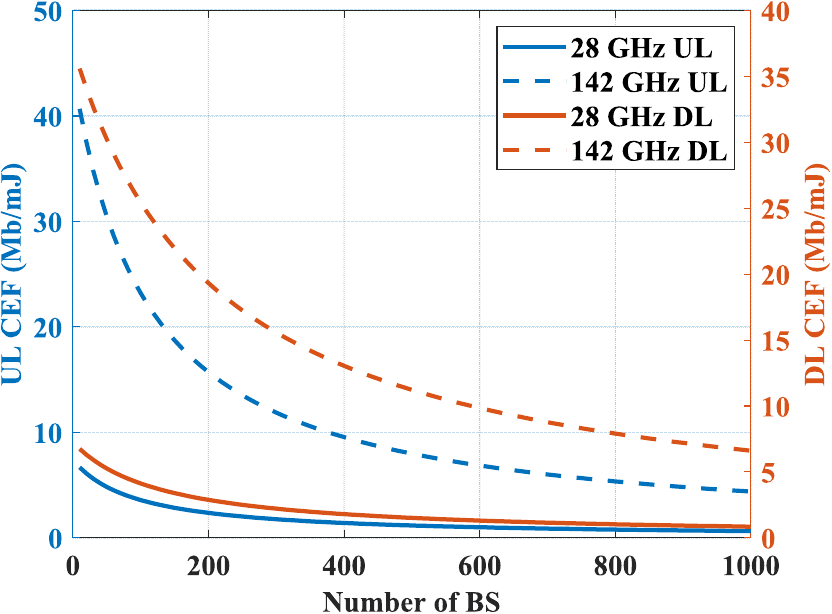}\label{figa6}}
	\vspace{-3pt}
	\caption{The impact of different phase shifter efficiency and different numbers of UE and BS on CEF}
	\vspace{-5pt}
\end{figure*}

\subsection{The Impact of Different Component Efficiency on CEF}

Fig. \ref{figa4}, considers variation in the phase shifters (PS) insertion loss as an example, and the results indicate sensitivity of CEF to the PS performance. Moreover, we observed that the changes in CEF with varying PS insertion loss were more pronounced in both the 28 GHz and 142 GHz uplink transmissions, while the changes in the downlink transmissions were relatively stable in subsequent regions. Furthermore, the CEF for both uplink and downlink transmissions at 142 GHz was higher than that at 28 GHz, implying that sub-THz systems require less RF power to achieve the same SNR as mmWave systems due to greater double directional antenna gains.
\vspace{-3pt}
\subsection{The Impact of Different Number of UE and BS on CEF}
\vspace{-3pt}
Expanding on the work in \cite{a10}, we explore the impact of varying UE and BS densification on CEF through simulation. We observed from Fig. \ref{figa5} that the 142 GHz scenario yields significantly larger uplink CEF, as compared to other scenarios, under varying UE numbers. Furthermore, with the increase in UE numbers, the rate of reduction in uplink CEF for 142 GHz is evidently lower than other scenarios, which indicates that sub THz communication exhibits good energy efficiency.

From Figure \ref{figa6}, the CEF trend for 28 GHz uplink and downlink transmissions is notably similar and significantly higher than that at 142 GHz, in line with findings in \cite{a10}. Additionally, an increase in BS density may enhance CEF during downlink transmissions at 142 GHz. These insights not only highlight strategies for optimizing transmission capacity across different frequency bands but also suggest AI/ML driven approaches can be employed to fine-tune parameters like BS densification. This can further enhance signal reliability, reduce interference, and inform spectrum allocation policies for optimal bandwidth utilization, ensuring system integrity.

\vspace{-3pt}
\section{Conclusions and Future Directions}
\vspace{-3pt}
In this study, we introduced the $W$, and showed its use in evaluating power efficiency across different wireless architectures. \(W\) when referred to the output of a cascaded system, is similar to \(F\), and offers a simple mathematical formulation for assessing wasted power. Such a formluation lends itself well to theory and to AI/ML applications for real-time control to improve power efficiencies in communication systems. Using \(W\), we assessed power efficiency of data centers through the PUE and compare them. Our analysis of a cascaded BS-UE communication system showed how system parameters impact \(W\) and data rate performance at 28 GHz and 142 GHz. We found that changes in phase shifter insertion loss affected CEF, especially in uplink transmissions. By examining different UE and BS densification, energy efficiency improved at 142 GHz, which can help in making better decisions for spectrum allocation and bandwidth use.

Looking ahead, as IoT devices become more common in wireless systems, there will be a greater need for energy-efficient solutions. RF energy harvesting at mmWave and sub-THz frequencies is an emerging area. With the growth of small cells and the introduction of Reconfigurable Intelligent Surface in 6G, designing energy-efficient systems will become even more important. \(W\) can be a useful tool in these designs. This paper provides a framework for understanding energy efficiency in communication systems. It offers valuable insights for improving next-generation communications and points to areas for future research to achieve greener communication methods particularly through AI/ML and in improving the power efficient designs of computing systems.

% \vspace{-3pt}
% \section{Acknowledgment}
% \vspace{-3pt}

% This work is supported by NYU WIRELESS industrial affiliates and NYU Tandon school of Engineering graduate fellowship.

% In conclusion, our research developed $W$ into a versatile framework for modern wireless systems. This study offered valuable metrics for optimizing energy efficiency in next-generation communication systems encompassing diverse network components such as data centers and transceivers. The study is aimed at contributing to the broader discourse on energy efficiency in communication systems.
% \vspace{-4pt}

\end{document}